\documentstyle{cupconf}
\begin{document}
\newcommand{\TEFF}{\mbox{T$_{eff}$}}
\newcommand{\TSTAR}{\mbox{T$_{\ast}$}}
\newcommand{\TE}{\mbox{T$_e$}}
\newcommand{\NE}{\mbox{n$_e$}}
\newcommand{\HII}{\mbox{H~II}}
\newcommand{\MSUN}{\mbox{M$_{\odot}$}}
\newcommand{\Halpha}{\mbox{H$_{\alpha}$}}
\newcommand{\OII}{\mbox{[OII]}}
\newcommand{\OIII}{\mbox{[OIII]}}
\newcommand{\SII}{\mbox{[SII]}}
\newcommand{\NII}{\mbox{[NII]}}
\newcommand{\Mgtw}{\mbox{Mg$_2$~}}
\newcommand{\Hbeta}{\mbox{H$\beta$~}}
\newcommand{\Fetw}{\mbox{Fe52~}}
\newcommand{\Feth}{\mbox{Fe53~}}
\newcommand{\Lsun}{\mbox{L$_{\odot}$}}
\newcommand{\Zsun}{\mbox{$Z_{\odot}~$}}
\newcommand{\Msun}{\mbox{M$_{\odot}$}}
\newcommand{\Zmin}{\mbox{$Z_{\rm m}~$}}
\newcommand{\Zmax}{\mbox{$Z_{\rm M}~$}}
\catcode`\@=11
\def\gsim{\ifmmode{\mathrel{\mathpalette\@versim>}}
    \else{$\mathrel{\mathpalette\@versim>}$}\fi}
\def\lsim{\ifmmode{\mathrel{\mathpalette\@versim<}}
    \else{$\mathrel{\mathpalette\@versim<}$}\fi}
\def\@versim#1#2{\lower 2.9truept \vbox{\baselineskip 0pt \lineskip 
    0.5truept \ialign{$\m@th#1\hfil##\hfil$\crcr#2\crcr\sim\crcr}}}
\catcode`\@=12
\def\boxit#1{\vbox{\hrule\hbox{\vrule\kern3pt\vbox{\kern3pt#1\kern3pt}\kern3pt
             \vrule}\hrule}}
\def\mg2{Mg$_2$}
\def\so{$\sigma_\circ$}
\def\spb{\smallskip\par\noindent$\bullet\;$}
\def\pn{\par\noindent}
\def\yr{hbox{\rm yr}}
\def\sss{\hbox{$\sigma_{\star}$}}
\def\me{\hbox{$M_{\rm e}$}}
\def\xe{\hbox{$X_{\rm e}$}}
\def\vto{\hbox{$V_{\rm TO}$}}
\def\ref{\par\noindent\hangindent=1truecm}
\def\feh{\hbox{{\rm [Fe/H]}}}
\def\MH{\hbox{\rm [M/H]}}
\def\mvhb{\hbox{$M_{\rm v}^{\rm HB}$}}
\def\lb{\hbox{$L_{\rm B}$}}
\def\lv{\hbox{$L_{\rm V}$}}
\def\lt{\hbox{$L_{\rm T}$}}
\def\hz{\hbox{$H_{\!\circ}$}}
\def\oz{\hbox{$\Omega_{\circ}$}}
\def\tgc{\hbox{$t_{\rm GC}$}}
\def\tgf{\hbox{$t_{\rm GF}$}}
\def\tz{\hbox{$t_{\circ}$}}
\def\ssss{\hbox{$\sigma_{\star}^2$}}
\def\mvto{\hbox{$M_{\rm V}^{\rm TO}$}}
\def\mr{\hbox{$M_{\rm r}$}}
\def\mto{\hbox{$M_{\rm TO}$}}
\def\mwd{\hbox{$M_{\rm WD}$}}
\def\mc{\hbox{$M_{\rm c}$}}
\def\fuv{\hbox{$F^{\rm UV}(Z)$}}
\def\fuvz{\hbox{$<\! F^{\rm UV}\! >_{\rm Z}$}}
\def\tjz{\hbox{$<\! t_{\rm j}\! >_{\rm Z}$}}
\def\tc{\hbox{$T_{\rm c}$}}
\def\te{\hbox{$T_{\rm e}$}}
\def\mef{\hbox{$M_{\rm H\!eF}$}}
\def\yms{\hbox{$Y_{\rm MS}$}}
\def\ttr{\hbox{$t_{\rm tr}\;$}}
\def\3/2{\hbox{${3\over 2}$}}
\def\rpn{\hbox{$R_{\rm PN}\;$}}
\def\zcrit{\hbox{$Z_{\rm crit}$}}
\def\vexp{\hbox{$v_{\rm exp}\;$}}
\def\mer{\hbox{$M_{\rm e}^{\rm R}$}}
\def\med{\hbox{$M_{\rm e}^{\rm D}$}}
\def\men{\hbox{$M_{\rm e}^{\rm N}$}}
\def\lsun{\hbox{$L_\odot$}}
\def\zsun{\hbox{$Z_\odot$}}
\def\msun{\hbox{$M_\odot$}}
\def\rsun{\hbox{$R_\odot$}}
\def\yr-1{\hbox{${\rm yr}^{-1}$}}
\def\mhr{\hbox{$M_{\rm H}$}}
\def\mh1{\hbox{$M_{\rm H}^1$}}
\def\mhf{\hbox{$M_{\rm H}^{\rm F}$}}
\def\mhuno{\hbox{$M_{\rm H}^1$}}
\def\mhs{\hbox{$M_{\rm H}^*$}}
\def\mhe{\hbox{$\Delta M_{\rm He}$}}
\def\mco{\hbox{$M_{\rm CO}$}}
\def\tpagb{\hbox{$t_{\rm P-AGB}$}}
\def\mhhb{\hbox{$M_{\rm H}^{\rm HB}$}}
\def\mrg{\hbox{$M_{\rm RG}$}}
\def\mhb{\hbox{$M_{\rm HB}$}}
\def\mrr{\hbox{$M_{\rm RR}$}}
\def\mvrr{\hbox{$M_{\rm V}^{\rm RR}$}}
\def\log{\hbox{${\rm Log\,}$}}
\def\mbol{\hbox{$M_{\rm bol}$}}
\def\dtp{\hbox{$\Delta t_{\rm peak}$}}\include{psfig}

\title[Origin of Bulges]
{Origin of Bulges}

\author[Renzini]
{Alvio Renzini\\ European Southern Observatory, \\ 
D-85748 Garching b. M\"unchen, Germany\\ arenzini@eso.org}

\maketitle

\begin{abstract}

Insight into the origin of bulges is saught in this review only from
the properties of their stellar populations. Evidence concerning the 
age of the Galactic bulge stellar population is reviewed first,
then the case of the bulge of M31 is discussed. The similarity of
bulges and ellipticals is then illustrated, inferring  that the
problems of the origin of bulges and of the origin of ellipticals may
well be one and the same: i.e. the origin of galactic spheroids. 
In this mood, the current evidence concerning the
age of the bulk stellar populations of early-type galaxies is then
reviewed, both for low- as well as high-redshift galaxies, and 
both for cluster as well as field ellipticals. All reported evidence 
argues for the bulk of stars in galactic spheroids having formed at
high redshift, with only minor late additions and small dependence on
environment. In the final, more speculative Section an 
attempt is made to evaluate how current 
formation scenarios can account for  this observational evidence.
The role of spheroids in the cosmic star formation and metal
enrichment history is also briefly discussed. Finally, some critical
questions are asked, which answers may help our further understanding 
of the formation and evolution of galactic spheroids.

\end{abstract}

\section{Introduction}
Much on our speculations on {\it how} bulges originated depend on what
we believe on {\it when} they formed. Some scenarios prefer bulges to
be young, or middle age, late comers anyway. Others prefer a rapid,
early build up of bulges, and push back to very
early times the epoch of their formation. For this reason I will
mostly concentrate on reviewing evidence on ages, leaving the
last section to speculations on origins. While they may provide
additional clues, some
morphological and dynamical properties -- such as bars, ripples,
 or peanut shapes -- are largely ignored in this review.

Section 2 focuses on the Galactic bulge, the one we can study best,
and in all details. Next closer bulge to us is that of M31, to which
Section 3 is dedicated. No other prominent bulge exists in the Local
Group (M33 does not really have a bulge), and Section 4 
emphasizes  that most bulges of spirals are quite similar to
ellipticals,
so the problem of bulge ages merges with  that of dating ellipticals,
and becomes the more general problem of dating {\it spheroids}. 
This is the subject
of Section 5, i.e. dating spheroids at low, as well as high redshifts.
In Section 6 cluster and field early-type galaxies are compared to
each other, and  
Section 7, on speculations, is last.

Overall, a wide body of observational evidences is presented showing
that the bulk of stellar populations in galactic spheroids are very
old. This is true all the way from the bulge of our own Galaxy to high
redshift cluster ellipticals. The main issue that remains open is
whether star formation and assembly of spheroids were concomitant
events, or whether the bulk of stars formed in smaller entities
that then hierarchically coalesced, with this process extending over much of
the cosmological time.

\begin{figure*}[htb]
\vspace {14cm}
\includegraphics{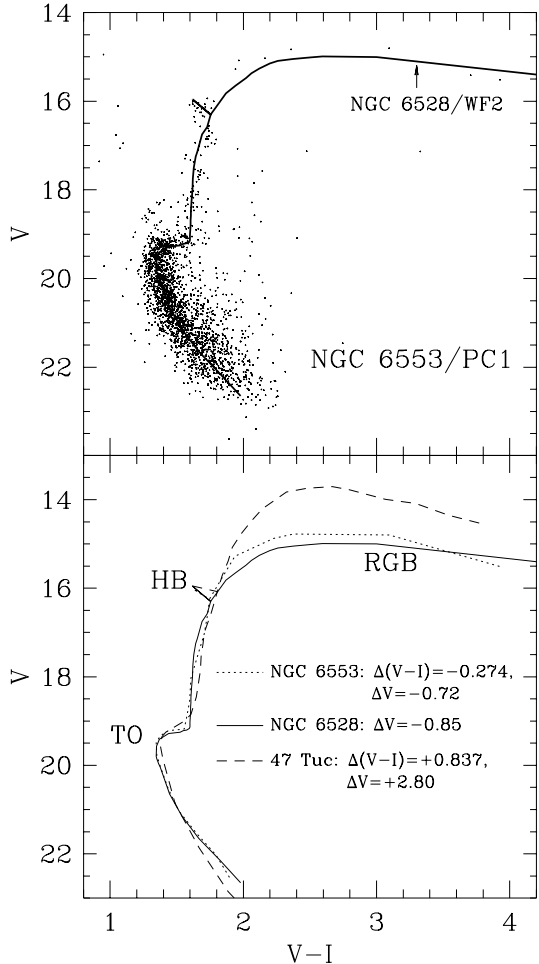}

\caption[]{Upper panel: the color-magnitude diagram of the bulge
globular NGC 6553 for stars in the PC field of WFPC2, with
superimposed the mean locus of the cluster NGC 5528, as sampled by
chip
\#2 of WFPC2. Lower panel: The mean loci of NGC NGC 6528, NGC 6553,
and 47 Tuc. Each locus has been shifted as indicated, in order to
bring into coincidence the end of the HB (from Paper II). 
} 
\end{figure*}

\section{The Age of the Galactic Bulge Relative to the Halo}
 
Dating of bulge stars is  complicated by several 
factors, such as crowding, depth effects, variable reddening,
metallicity dispersion, and contamination by foreground disk stars. 
In an attempt to circumvent some  of these limitations 
Ortolani et al. (1995)
have selected the bulge 
globular clusters NGC 6528 and NGC 6553 for HST study. These clusters are 
respectively 
located at $\sim 4^\circ$ and $6^\circ$ from the galactic center, and their 
overall metallicity [M/Fe] is about solar (Barbuy et
al.  1999), close to the average for stars 
in Baade's Window (McWilliam \& Rich 1994). Like most other clusters within
$\sim 3$ kpc from the Galactic center, they belong to the population
of Bulge globular clusters, having the same kinematical properties and
metallicity distribution of Bulge stars (e.g. Minniti 1995). (To
qualify these clusters as {\it disk clusters} is clearly a misnomer.)

The upper panel in Fig. 1 shows the CMD of NGC 6553 as sampled by the
PC1 chip of WFPC2, that combines good statistics with relatively low
differential reddening. Superimposed on it is the mean locus of the
CMD of NGC 6528 as sampled by the WF2 chip of WFPC2. The data points
of NGC 6553 have been dereddened as indicated in the lower panel so to
make its turnoff color equal to that of NGC 6528. Then the CMD of this
latter cluster has been shifted vertically to bring its horizontal
branch (HB) to coincide with that of NGC 6553.
Note that the mean
locus of NGC 6528 provides an excellent fit to the NGC 6553 data, from
the MS all the way to the tip of the RGB. The virtual identity of the
CMD of the two clusters is further demonstrated in the lower panel of
Fig. 1, where the men loci of the two clusters are compared to each
other.

Also shown in the lower panel of Fig. 1 is the mean locus of the inner
halo globular cluster 47 Tuc ([Fe/H]=--0.7), which has been shifted in
color and magnitude in order to bring into coincidence its 
HB with that of the two bulge clusters. As can be seen in
this figure, the luminosity difference between the HB and the main
sequence
turnoff of the the Bulge clusters 
is the same (or even slightly larger) of that of  47 Tuc. 
This comparison demonstrates that the two Bulge
clusters are as old as the halo clusters (to within $\pm\sim 2$ Gyr), and
therefore the bulge underwent rapid chemical enrichment to
solar abundance and beyond, very early in the evolution of our Galaxy.
Due to the {\it relative} nature of the dating procedure, this
conclusion
is independent of uncertainties in reddening, distance, and {\it
absolute} age determinations.

The next step in the Ortolani et al. study is represented by an
attempt
at dating the Bulge {\it field} stellar population itself, still in a
relative fashion with respect to the clusters.
Fig. 2 shows that the MS luminosity function of the
cluster NGC 6528 is indistinguishable from that of the stars in Baade's
Window
(the low-reddening bulge field at $\sim 4{\deg}$ from the Galaxy
center), that was  obtained from observations with the ESO NTT with
superb seeing
($0''.4$). From this comparison Ortolani et al. 
 infer that the whole Bulge formed
quickly, some 15 Gyr ago (if this is the age of the halo clusters), 
and set an upper limit of
$\sim 10\%$ by number to any intermediate age population in the Bulge.
Indeed, a larger proportion of intermediate age stars would have
resulted in a shallower fall off of the bulge luminosity function
around TO (i.e., for $20.5\gsim V\gsim 19.5$), where instead it
coincides with that of the cluster.

\begin{figure*}[htb]
\vspace {10cm}
\includegraphics{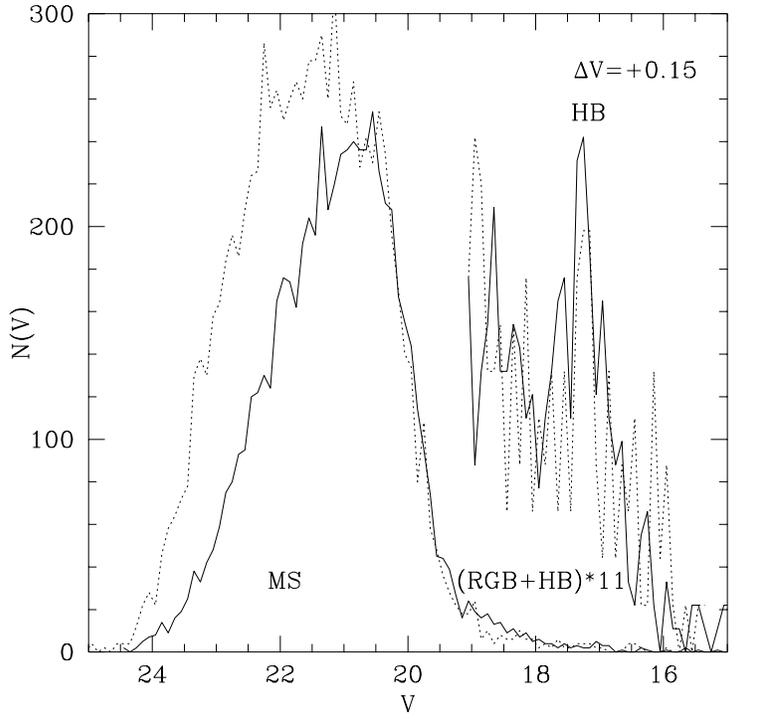}
\caption[]{The luminosity function (LF) of main sequence (MS) and red giant 
(RGB+HB) 
stars  in NGC 6528 (WF2 field,
dotted line) and in Baade's Window (BW, solid lines). 
The cluster LF has been shifted by $\Delta V=+0.15$ so as
to bring into coincidence its HB peak (marked on the figure) with that of
BW, and multiplied by a factor of 2 so as to normalize the two
distributions at 
$V=20.45$, where both are reasonably complete, or to the same number of 
RGB+HB stars brighter than V=19.45 in this figure. Note that below $V\simeq 21$
the bulge LF is progressively more incomplete compared to that of the cluster.
The cluster LF has been suitably
broadened with a
Montecarlo simulation to mimic the depth effect present in the BW
field.
For this display, the RGB+HB LFs have been multiplied by a factor 11,
in order to avoid overlap with the LF
of the disk foreground stars (from Ortolani et al. 1995).}
\end{figure*}

Further insight on the formation time scale of the Bulge comes from
the detailed abundance studies of Bulge stars. In 12 field K giants
McWilliam \& Rich (1994) find a moderate $\alpha$-element enhancement
([Mg/Fe]$\simeq$[Ti/Fe]$\simeq$+0.3, but with
[Ca/Fe]$\simeq$[Si/Fe]$\simeq 0$), moderate r-process element
enhancement, while s-process elements appear solar with respect to
iron.  Barbuy et al. (1999) have analyzed 2 stars in NGC 6553 finding
somewhat more enhanced $\alpha$-element overabundance, with
[Na/Fe]$\simeq$[Al/Fe]$\simeq$[Ti/Fe]$\simeq$+0.6 and 
[O/Fe]$\simeq$[Mg/Fe]$\simeq$[Si/Fe]$\simeq$+0.3. Note that different 
systematcs may go a long way towards explaining the differences
between these two studies. General consensus exists on the
interpretation
of $\alpha$-element and r-process element enhancements as due to a
{\it quick} star formation and metal enrichment, 
with elements produced by Type II supernovae being
incorporated into new stars before the bulk of iron from Type Ia SNs
is produced. However, how quick is {\it quick} remains
uncertain. Basically,
the bulk of stars must form before the explosion of most SNTa's, but
the actual {\it distribution} of SNIa explosion times following a
burst of star formation remains empirically indetermined and
theoretically very model dependent (cf. Greggio 1996). According to
general wisdom it takes at least $\sim 1$ Gyr for a fair fraction of
SNIa to release their iron. If so, at least 90\% of the Bulge stars
formed within  the first Gyr of the object that we now call the Milky Way.
 In conclusion, the
{\it fossil} evidence tells us that the whole Galactic spheroid is
pretty old indeed, and formed on a rather short timescale.

There are important lessons to draw from these conclusions.  Our Milky
Way is a rather late-type spiral galaxy in a very loose group that is
located rather away from major density peaks in the distribution of
galaxies. Nevertheless, her whole spheroidal component looks $\sim$
one Hubble time old, from the halo globular clusters all the way to
the inner bulge. With a mass of $\sim 2\times 10^{10}\msun$, the old
age for the bulk of the spheroidal population implies an average star
formation rate $\sim 20\,\msun \yr-1$ at the epoch of spheroid
formation, some 14-15 Gyr ago (having assumed $\sim 10^9$ yr for the
duration of the star formation process). This value is as small as the
smallest star formation rates 
of $z\gsim 3$ galaxies (Steidel et al. 1998). Such
galaxies have also effective radii of 1-3 kpc (typical of galactic
bulges, cf.  Giavalisco et al. 1996), and it is rather tempting to
speculate that with Lyman-break galaxies one may have caught bulge
formation in the action. With the Galactic spheroid accounting for $\sim
20\%$ of the stellar mass of the Milky Way, one can conclude that
$\gsim 20\%$ of all stars in our Galaxy have formed ``at fairly high
redshift''.

\section{The Next Bulge: M31}

In ground based and pre-COSTAR HST studies the suspicion had been
advanced for the presence in the bulge of M31 of a major
intermediate-age component, as suggested by the detection of putative
bright AGB stars (e.g. Rich \& Mould 1991; Rich, Mould, \& Graham 1993; Rich \&
Mighell 1995; Davidge et al. 1997).  However, bright AGB stars
($M_{\rm bol}\simeq -5$) are also produced by old, metal rich globular
clusters, such as the Bulge clusters discussed in the previous section
(Frogel \& Elias 1988; Guarnieri, Renzini, \& Ortolani 1997). Moreover, with
insufficient angular resolution blends of RGB stars can be mistaken
for bright AGB stars (e.g. Renzini 1998b), and the presence of an
intermediate age population in the bulge of M31 could not be
unquestionably proven with such data.

\begin{figure*}[htb]
\vspace {10cm}
\includegraphics{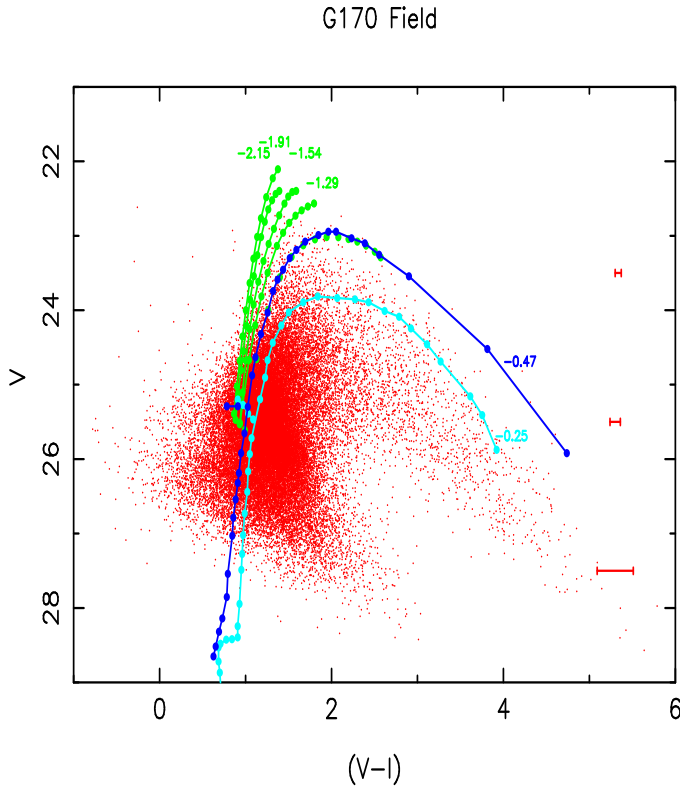}
\caption[]{The WFPC2 color-magnitude diagram of the field near the
globular cluster G170, located at a projected distance of 
$\sim 1.55$ kpc from the center of the
bulge (from Jablonka et al. 1999). Also shown is the red giant branch
loci for several metallicities [Fe/H], as indicated.
}
\end{figure*}

WFPC2 observations of the bulge of M31 are now becoming
available. Jablonka et al. (1999) have obtained deep CMDs for various
locations in the bulge of M31, confirming that what on low resolution
images appeared as {\it bright AGB stars} are indeed blends of fainter
stars. Fig. 3 shows one of such CMDs, relative to the field in the
vicinity of the very metal rich globular cluster G170, located at a
projected distance of 1.55 kpc from the center of M31. The CMD of the
field near the cluster G177 (at 0.8 kpc from the center) is virtually
identical. Perhaps the most stryking aspect of this CMD is the
predominance of a fairly homogeneous metal rich population, with the
upper RGB bending down in the $V-(V-I)$ CMD due to strong TiO
blanketing as typical of metal rich globular clusters (see Fig. 1).
Very few metal poor star appear to be present, while the bulk of stars
are more metal rich than [Fe/H]=--0.25.

In conclusion, there is no evidence for an intermediate age population
in the bulge of M31. Its almost uniformly metal rich population 
points to the presence of a ``G-dwarf Problem'', which may
be a general characteristics of (at least) the inner regions of galactic
spheroids (e.g. Greggio 1997).  The metallicity distribution of the
M31 bulge may provide important insight for understanding the
formation process. It rises two intriguing questions: 1) Where are the
stars that produced the metals now locked in the bulge stars we see?
and 2) Where have the metals produced by this bulge sellar population
gone? The tentative answer to the first question is ``they are out in the
halo of M31'', which could be tested extending deep HST imaging to larger
galactocentric distances (but see Rich, Mighell, \& Neill 1996). 
The tentative answer to the second question 
is ``they have
been ejected out in the IGM by an early galactic wind''. 
If these are the
correct answers, then the even more tentative conclusion is that the bulge
formed outside-in by dissipative merging and collapse of mostly
gaseous pregalactic lumps. with the resulting starburst then ejecting
the residual gas and a lot of metals along with it.

\section{Bulges vs Ellipticals}
The properties of bulges are extensively reviewed at this meeting, and
there is no point trying to summarize them here. In this section I
would like to emphasize only one aspect: the close similarity of the
bulges of spiral galaxies with elliptical galaxies. While also this
aspect is further illustrated by others at this meeting, Fig. 4 gives
a very direct impression of the extent to which bulges are similar to
ellipticals (from Jablonka, Martin, \& Arimoto 1996), and therefore
may share a common origin. The bulk of bulges appear to follow
precisely the same Mg$_2-M_{\rm r}$ relation of ellipticals, with just
a minority of them (i.e. 5 out of 26 in the Jablonka et al. sample)
having Mg$_2$ values appreciably lower than those of ellipticals of
similar luminosity. The same similarity also exists between the
Mg$_2-\sigma$ relations of bulges and ellipticals (Jablonka et
al. 1996).

\begin{figure*}[htb]
\vspace {10cm}
\includegraphics{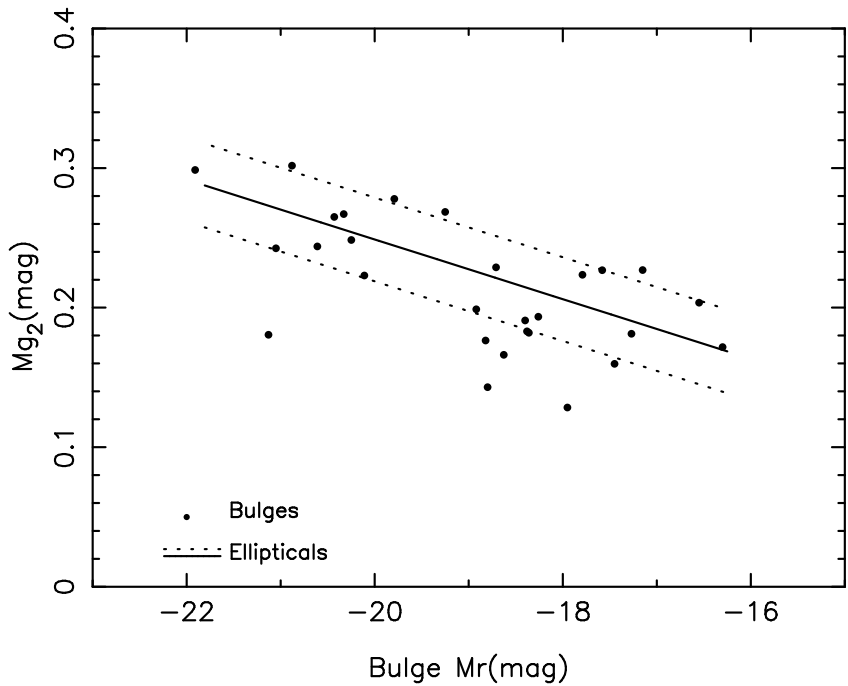}
\caption[]{The Mg$_2-M_{\rm r}$ relation for a sample of bulges. The
solid
line is the mean relation for elliptical galaxies, and the dotted
lines limit the area occupied by ellipticals (from Jablonka et al. 1996). 
}
\end{figure*}

As well known, the Mg$_2$ index depends on both age and metallicity;
actually on both the age and metallicity {\it
distributions}. Therefore,
the close similarity of the Mg$_2-M_{\rm r}$ relations argues for
spiral
bulges and ellipticals sharing a similar star formation
history and chemical enrichment. One may argue that origin and
evolution have been very different, but differences in age
distribution are precisely compensated by differences in the
metallicity distributions. This may be difficult to disprove, and I
tend
to reject this alternative on aesthetic grounds. It requires an
unattractive cosmic
conspiracy, and I would rather leave to others the burden of defending
such a scenario.

In conclusion, it appears legitimate to look at bulges as ellipticals
that happen to have a prominent disk around them, or to ellipticals as
bulges that for some reason have missed the opportunity to acquire
or maintain a prominent disk. Therefore, we can legitimately refer to 
{\it spheroids} as the class of objects that includes ellipticals and
the bulge+halo component of spirals. In this mood, the problem of the
origin of bulges becomes the problem of the origin of spheroids.

\section{The Epoch of Spheroid Formation}
Great progress has been made in recent years towards charting and
modeling galaxy formation and evolution. Yet, the origin of the galaxy
morphologies, as illustrated by the Hubble sequence, has so far defied
a generally accepted explanation. This is also the case for
spheroids, i.e. bulges and ellipticals alike, with two quite different
scenarios still confronting each other. In one scenario spheroids come
from the destruction of pre-existing disks or part of them. In the
case of ellipticals, by merging spirals, a widely entertained notion
since the original proposal by Toomre (1977). In the case of bulges,
by some bar instability randomizing the orbits of stars originally in
the inner part of a disk (e.g. Combes et al. 1990; Raha et al. 1991;
Hasan, Pfenninger, \& Norman 1993), or by being merger 
remnant ellipticals that
managed to re-acquire a new disk.  This latter scenario is now
motivated by hierarchical clustering cosmologies, and ellipticals are
modeled to form through a series of merging events (between spirals)
taking place over a major fraction of the cosmological time
(e.g. Baugh, Cole, \& Frenk 1996; Kauffmann 1996).

The other scenario assumes instead the whole baryonic mass of the galaxy
being already assembled at early times in gaseous form, and for this
reason it is sometimes qualified as {\it monolithic}. The original
idea can be traced back to the Milky Way collapse model of
Eggen, Lynden-Bell, \& Sandage (1962), with early 
examples including the models of Larson (1974) and Arimoto \& Yoshii (1987). 
In this case, the disk of spirals is a late comer, somehow acquired
later by a more ancient spheroid.

\begin{figure*}[htb]
\vspace {9.3cm}
\includegraphics{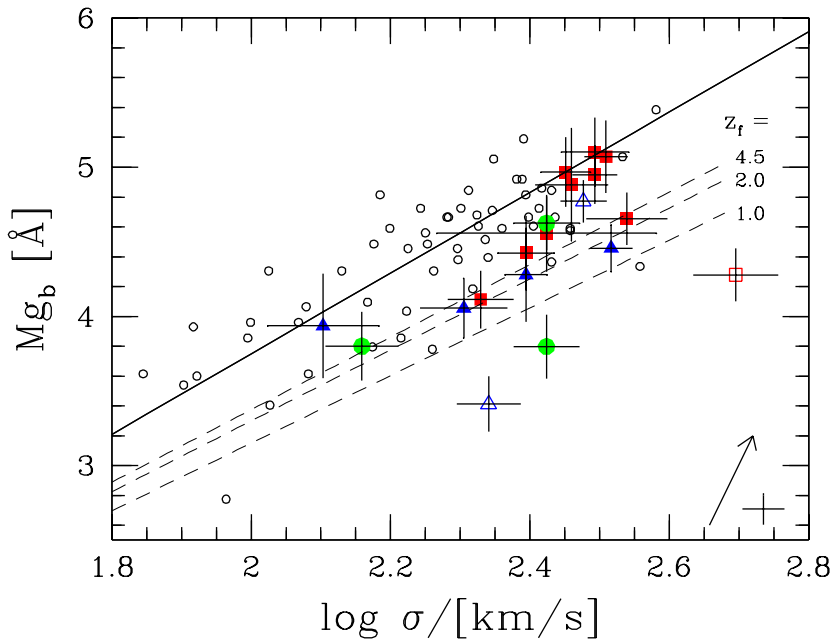}
\caption[]{The Mg$_{\rm b}-\sigma$ relation for a sample of
ellipticals in two clusters at $z\simeq 0.37$ (symbols with error
bars) is compared to the same relation for a sample of galaxies 
in the Virgo and Coma clusters (from Bender et al. 1997). The dashed
lines represent the expected location of single burst, passively
evolving galaxies for various formation redshifts (with $H_\circ=50$, 
$q_\circ=0.5$). The aperture correction is shown near the lower/right corner.
}
\end{figure*}

Through the 1980's much of the debate focused on the age of
ellipticals as derived from the integrated spectrum of their stellar
populations. In general, advocates of the merger model favored an
intermediate age for the bulk of stars in ellipticals, but the matter
remained controversial given the well know age-metallicity degeneracy and
the crudeness of stellar population models of the time
(for opposite views see O'Connell 1986, and Renzini 1986). 

\begin{figure*}[htb]
\vspace {14cm}
\includegraphics{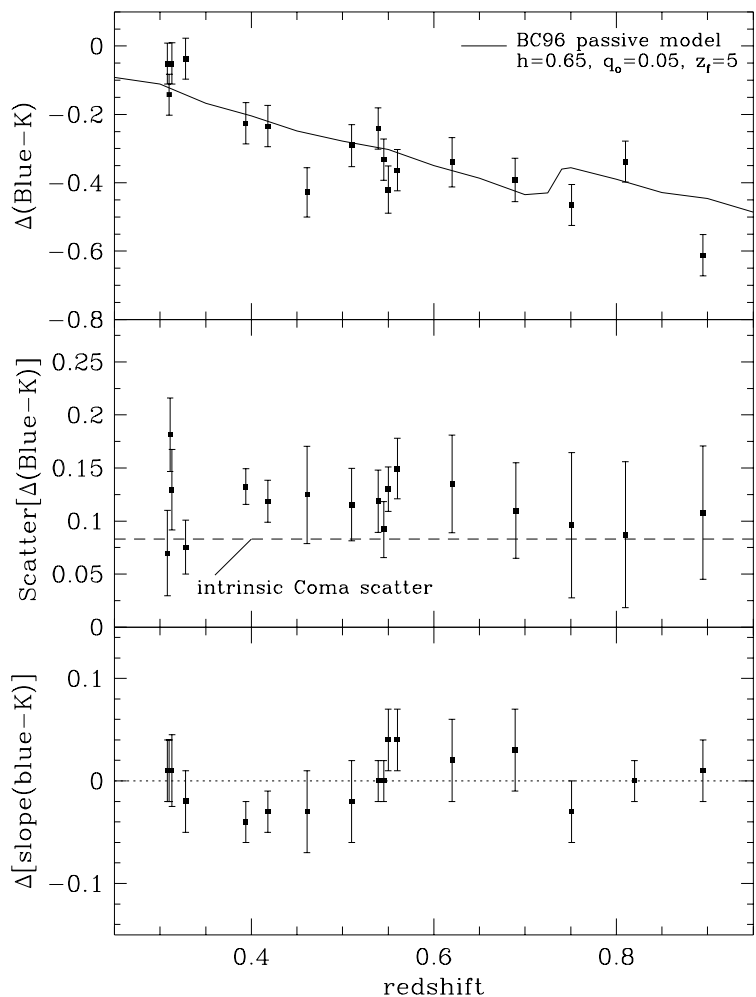}
\caption[]{The color evolution of early-type galaxies in clusters out
to $z\simeq 0.9$ (Stanford, Eisenhardt, \& Dickinson 1997; Dickinson
1997). The ``blue'' band is tuned for each cluster to approximately
sample the rest frame $U$-band, while the $K$ band is always in the
observed frame. Top panel: the redshift evolution of the blue$-K$
color relative to the Coma cluster. A purely passive evolution models
is also shown.  Middle panel: the intrinsic color scatter, having
removed the mean slope of the color-magnitude relation in each cluster
and the contribution of photometric errors. The intrinsic scatter of
Coma galaxies is shown for reference. Bottom panel: the redshift
evolution of the slope of the (blue$-K)-K$ color-mag diagram, modulo
the slope for galaxies in Coma.  }
\end{figure*}

A first breakthrough came from noting the tightness of the
color-$\sigma$ relation of ellipticals in the Virgo and Coma
clusters (Bower, Lucey, \& Ellis 1992). This demands a high degree of
{\it synchronicity} in the star formation history of ellipticals, that
is most naturally accounted for by pushing back to early times most of
the star formation. Making minimal use of stellar 
population models,  this approach provided for the first time
a {\it robust} demostration that at least
{\it cluster} ellipticals are made of very old stars, with the bulk
of them having formed at $z\gsim 2$. 

The main lines of the Bower et al. argument are as follows. 
The observed color scatter of cluster ellipticals is related to
the age dispersion among them by the relation:
\begin{equation}
\delta (U-V) = {\partial (U-V)\over\partial t}(t_{\rm H}-t_{\rm
F})
\end{equation}
where $t_{\rm H}$ and $t_{\rm F}$ are  the age of the ``oldest''
and ``youngest'' galaxies, respectively. Here by age one intends the 
luminosity-weighted age of the stellar populations that constitute such
galaxies. The time derivative of the color is obtained from evolutionary
population synthesis models, which give $\partial (U-V)/\partial
t\simeq 0.02$ mag/Gyr for $t\simeq 10$. The observed color scatter is
$\delta (U-V)\simeq 0.04$ mag, consistent with pure
observational errors. Hence, one gets $t_{\rm H}-t_{\rm F}\lsim 0.04/0.02=2$
Gyr, and if the oldest galaxies are 15 Gyr old, the youngest ones
ought to be older than 13 Gyr, from which Bower et al. conclude they
 had to form at $z\gsim 2$. If the oldest galaxies were instead  as young as, say 5
Gyr, then the youngest should be older than at least 3 Gyr, which
would require a high degree of synchronicity in their formation, which
seems unlikely.

Evidence in support of the Bower et al.
conclusion has greatly expanded through the 1990's, 
and is now compelling. 
This came from the tightness of the fundamental plane
relation for ellipticals in local clusters (Renzini \& Ciotti 1993),
from the tightness of the color-magnitude relation for ellipticals in
clusters up to $z\sim 1$ (e.g., Aragon-Salamanca et al. 1993; Taylor
et al. 1998; Kodama et al. 1998; Stanford,
Eisenhardt, \& Dickinson 1998), and from the
modest shift with increasing redshift in the zero-point of the fundamental
plane, Mg$-\sigma$, and color-magnitude relations of cluster
ellipticals (e.g., Bender et al. 1997;
Dickinson 1995; Ellis et al. 1997; van Dokkum et al. 1998;
Pahre, Djorgovski, \& de Carvalho 1997; Stanford,
Eisenhardt, \& Dickinson 1998; Kodama et al. 1998). All these studies
agree in concluding that most stars in ellipticals formed at $z\gsim
3$, though the precise value depends on the adopted cosmology. 
Fig. 5 illustrates the case of the Mg$-\sigma$ relation for
ellipticals in two clusters at $z\simeq 0.37$, while Fig. 6 documents
the constancy of the color disperion of cluster ellipticals all the
way to $z\sim 1$.

It is worth emphasizing that all these studies follow the 
methodological approach pioneered by Bower et al. (1992). They focus indeed on
the tightness of  some
correlation amomg the global properties of cluster ellipticals, which
sets a robust constraint on their age dispersion as
opposed to an attempt to
date individual galaxies. Moreover, the move to high redshift offers two
fundamental advantages. The first advantage is that looking at high
$z$ provides the best
possible way (I should say {\it the} way) of removing the
age-metallicity degeneracy. If spheroids are made of intermediate-age,
metal rich stars, they should become rapidly bluer and then disappear
already at
moderate redshift (e.g. Kodama \& Arimoto 1997). The
observational opportunity of studing galaxies at large lookback times 
makes quite obsolete attempts at finding combinations of spectral indeces
that may distinguish between age and metallicity effects in nearby
galaxies. The second
advantage is that at high redshift one gains more {\it leverage}: for given
dispersion in some observable one can set tighter and tighter limits
to the age dispersion. This comes from the color time derivatives 
being larger the younger the population. For example, the derivative 
$\partial (U-V)/\partial t$ is $\sim 7$ times larger at $t=2.5$ Gyr
than it is at $t=12.5$ Gyr (e.g. Maraston 1998), and therefore a given
dispersion in this rest-frame color translates into a $\sim 7$ times
tighter constraint on age and therefore on formation redshift.
This is further illustrated also by the case of isolated high redshift
ellipticals. For example, Spinrad et al. (1997) found a {\it fossil}
(i.e. passively evolving) elliptical at $z=1.55$ for which they infer
an age of at least 3.5 Gyr, hence a formation redshift in excess of
$\sim 5$. An even much higher formation redshift may be appropriate
for the extremely red galaxy in the NICMOS field of the
HDF-South, which spectral energy distribution is best accounted for by
an old, passively evolving population at $z\simeq 2$ (Stiavelli et al. 1999). 
\begin{figure*}[htb]
\vspace {14cm}
\includegraphics{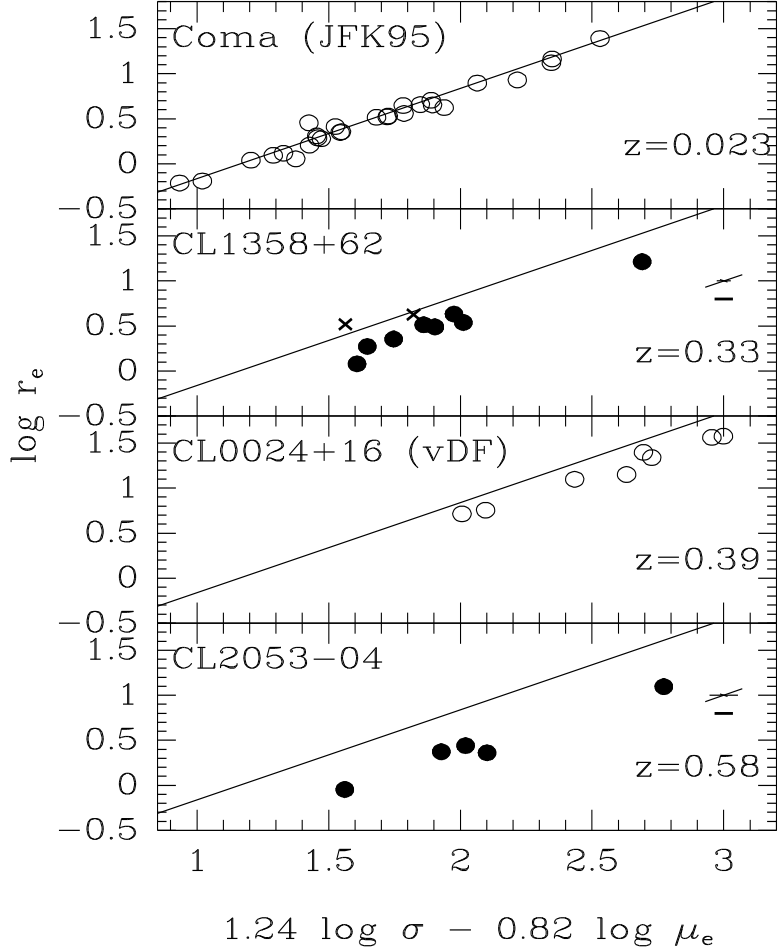}
\caption[]{The fundamental plane relations of clusters at increasing
redshifts (Franx et al. 1997). Note that the slope of the fundamental
plane remains constant. This is also the case when including the
cluster MS1054--03 at $z=0.83$ (van Dokkum et al. 1998). }
\end{figure*}

\section{Cluster vs Field Spheroids}

Much of the evidence discussed in the previous Section is restricted
to cluster ellipticals.  In hierarchical models, clusters form out of
the highest peaks in the primordial density fluctuations, and cluster
ellipticals completing most of their star formation at high redshifts
could be accommodated in the model (e.g. Kauffmann 1996; Kauffmann \&
Charlot 1998a). However, in lower density, {\it field} environments,
both star formation and merging are appreciably delayed to later times
(Kauffmann 1996), which offers the opportunity for an observational
test of the hierarchical merger paradigm.

The notion of field ellipticals being a less homogeneous family compared
to their cluster counterparts has been widely entertained, though the
direct evidence has been only rarely discussed.  Visvanathan \& Sandage
(1977) found cluster and field ellipticals to follow the same
color-magnitude relation, but Larson, Tinsley, \& Caldwell (1980) --
using the same database -- concluded that the scatter about the mean
relation is larger in the field than in clusters.
More recently, a larger scatter in field versus
cluster ellipticals was also found for the fundamental plane 
relations by de Carvalho \& Djorgovski (1992). However,
at least part of the larger scatter among field ellipticals certainly comes
from their distances being more uncertain than for clusters.

Taking advantage of a large sample ($\sim 1000$) of early-type
galaxies with homogenously determined Mg$_2$ index and central
velocity dispersion, Bernardi et al. (1998) have recently compared the
Mg$_2-\sigma$ relations (which are distance independent!) of cluster
and field galaxies, and the result is shown in Fig. 8.  As it is evident
from the figure, field, group, and cluster ellipticals all follow
basically the same relation. The zero-point offset between cluster and
field galaxies is $0.007\pm 0.002$ mag, with field galaxies having
lower values of \mg2, a statistically significant, yet very small
difference.  This is in excellent agreement with the offset of
$0.009\pm 0.002$ mag, obtained by J\/orgensen (1997) using 100 field
and 143 cluster galaxies.

Using the time derivative of the Mg$_2$ index from synthetic stellar 
populations, Bernardi et al. conclude that the age difference between
the
stellar populations of cluster and field early-type galaxies is at
most $\sim 1$ Gyr. The actual difference in the mass-weighted age
(as opposed to the luminosity-weighted age) could be significantly
smaller that this. It suffices  that a few galaxies have undergone a minor star
formation event some Gyr ago, with this having taken place
preferentially among field galaxies.

\begin{figure*}[htb]
\vspace {14cm}
\includegraphics{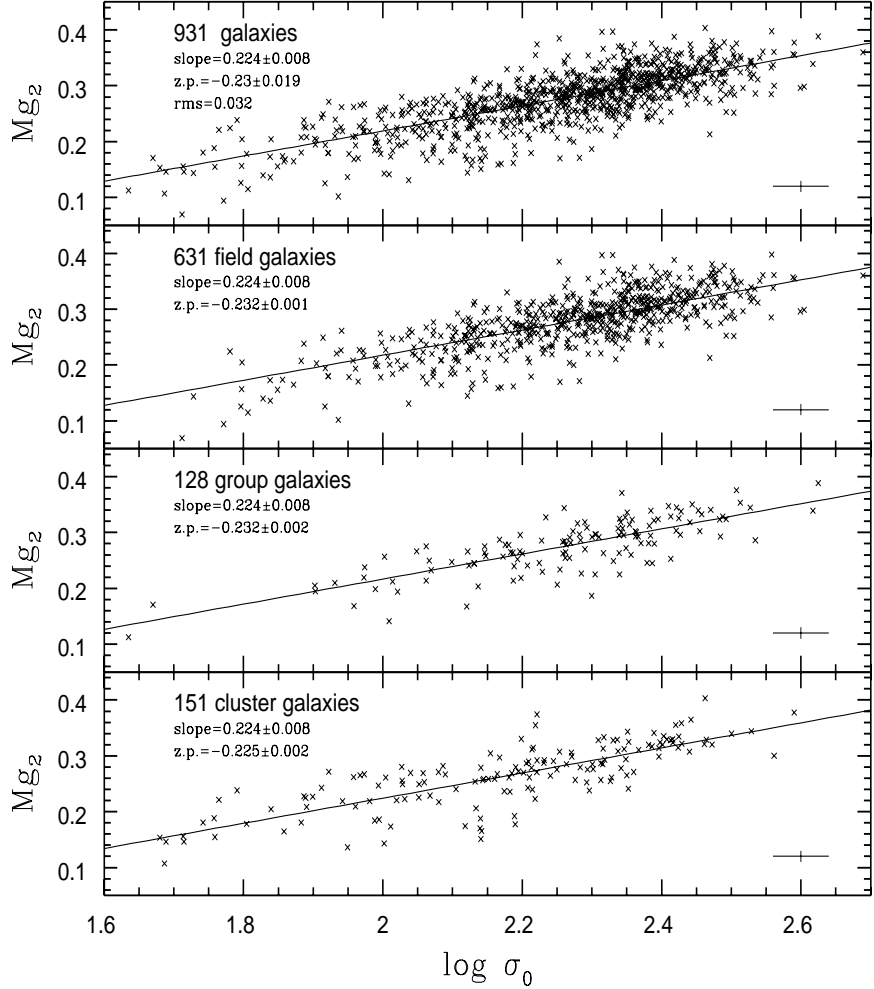}
\caption[]{The \mg2--$\sigma$ relation for a sample of early-type
galaxies (upper panel), as well as for the field, group and cluster
subsamples (lower panels), from Bernardi et al. (1998). 
The corresponding number of objects,
the slope, and the zero-point (z.p.) are shown in the upper/left
corner of 
each panel. The least squares fits to the \mg2--\so\ relation
are also shown as solid lines. For the three subsamples the slope as
derived for the total sample was retained, and only the zero-point was
determined. The typical error bar is shown in the lower/right corner.
}
\end{figure*}

The comparison between these empirical findings and the theoretical
simulations is somehow complicated by the rather loose way in which
cluster, group, and field environment are defined in the observational
studies on the one hand, and in the theoretical simulations on the
other. For example, in the models of Kauffmann (1996) there is a $\sim
4$ Gyr age difference between model ellipticals now residing in a
$10^{15}\msun$ dark matter halo and those residing in a
$10^{12}-10^{13}\msun$ halo.  This age difference
would correspond to a difference $\Delta$\mg2$\simeq 0.023$ mag, which
Bernardi et al. data exclude at the $\sim 4.5\sigma$ level. However, in
Kauffmann \& Charlot (1998a) cluster and field are defined as those
dark matter halos with circular velocity $V_{\rm c}=1000$ km s$^{-1}$
and $<600$ km s$^{-1}$, respectively, and the age difference between
model ellipticals in such two environments is greatly
reduced. Empirically, Bernardi et al. have assigned to the field those
galaxies that do not belong to known clusters or groups, and one does
not know what
the average circular velocity is in such environments.
Clearly, the problem is to find a common language between
observers and model makers, i.e. a common way of defining cluster and
field environments before comparing to each other data and
simulations.

\section{Discussion}

As documented in the previous sections, compelling evidence now
exists for the bulk of stars in galactic spheroids being
very old, i.e. formed at redshifts beyond $\sim 3$, and possibly even
much beyond this value. This applies to 
ellipticals and bulges alike, in clusters as well as in the lower
desity regions still inhabited by spheroids, including the bulge of
our own Galaxy. This is what was expected
(actually postulated) in the monolithic collapse scenario, while it 
appears to be quite
at variance with most realizations of the hierarchical merging scenario.

\subsection{Monolithic vs Hierarchical}

The fact that spheroids are made of old stars does not necessarily 
invalidate the hierarchical merging
paradigm, which actually offers a still unique description on how
large galaxies could have been assembled. One possibility to comply with
with the observations may be to tune
hierarchical models to mimic the monolithic model, 
by pushing most of the action back to an earlier cosmological
epoch. With most of the merging taking place at high
redshifts, among still mostly gaseous components, 
merging itself would promotes widespread starburst activity. The natural 
observational counterparts of these events may be represented by the
Lyman-break galaxies at $z\gsim 3$ (Steidel et al. 1996), where star
formation rates could be as high as $\sim 1000\; \msun\yr-1$ 
(Dickinson 1998). It remains to be explored whether such tuning of algorithms
and parameters  of the hierarchical model could produce model
universes fulfilling all other observational constraints.
Alternatively, stars now in spheroids
do indeed form at very high redshifts, but they are assembled into big
spheroids only at much later times (as favored, e.g. by Kauffman 1996).

One testable prediction of the hierarchical merging model is that --
obviously -- bigger galaxies form later by assembling smaller pieces,
and their stellar populations are appreciably younger than those of smaller
galaxies. Therefore, intrinsically brighter galaxies should get bluer at
a faster rate with increasing $z$, compared to fainter ones.
As a consequence, the color-magnitude, color-$\sigma$, Mg-$\sigma$
and fundamental plane relations should {\it flatten} with increasing redshift
(lookback time). No such effect has been detected yet: the slope of
the color-magnitude relation appears to be the same all the way to at
least $z\simeq 1$ (see bottom panel in Fig. 6). The predicted
 flattening is actually a
consequence of the postulate that ellipticals are made by merging
spirals, with the  gas in the disks being turned into stars when the
two dark matter halos merge. Hence, in this frame 
late merging implies late star formation as well. On the other hand,
it remains to be seen whether dissipationless merging of gas-free galaxies 
can produce the spheroids we see at low redshift, with their very high
phase-space density. If so, the color-mag and similar relations should
actually get {\it steeper} with increasing redshift.

The other prediction of the hierarchical model is that big galaxies
should progressively disappear with increasing redshift, and several
claims have been made pros and cons the actual disappearence of
elliptical in various redshift surveys. Unfortunately, this approach is less
conclusive than it may appear at first sight: when ellipticals are
selected following either color or morphological criteria a small
residual star formation should suffice to let otherwise old galaxies to
drop out of the selected samples, even if their main (spheroidal)
body is already in place.

To overcome the intrinsic weakness of this approach,  Kauffmann \&
Charlot (1998b) avoid using either color or morphology criteria, and
adopt a pure $K-$band magnitude limited selection criterion. In this
way the number evolution of massive galaxies is followed,
independently of morphology or trace star formation, hence providing
a more fundamental test of the models. Comparing to a $K<19$ sample
of galaxies with measured redshift, they conclude that their pure luminosity
evolution (PLE) models
are excluded by a large margin. Such models would predict $\sim 50\%$
of the galaxies in the sample to be at $z>1$, while only $\sim 10\%$
is observed, hence they argue for number evolution due to merging
being at work. The same test can be attempted on the somewhat bigger 
$K-$band magnitude limited sample of Cohen et al. (1998), which includes
195 objects down to $K=20$. Among these objects, 24 turned out to be
stars and for 34 objects no redshift could be determined. Among the
residual 137 objects, 21 have $z>1$. The vast majority of objects without
a measured redshift are likely to be galaxies at $z>1$, whose strong
spectral features have moved out of the range of the optical spectrograph.
If so, the sample would have $\sim 21+34=55$ out $\sim 137+34=171$
galaxies at $z>1$, or $\sim 32\%$. Interpolating on Figure 4 in
Kauffmann \& Charlot (1998b) one can roughly estimate that their PLE 
model predicts $\sim 60\%$ of galaxies in a $K\le 20$ sample to be at
$z>1$, while their hierarchical model predicts $\sim 10\%$. So, the
Cohen et al. sample suggests a value that is just midway between the 
predictions of the two  models. Clearly, existing samples are
still too small for reaching any firm conclusion, 
especially when considering that
large fluctuations may take place between one pencil beam survey and another
due to fluctuations in the sampled large scale structures. For
example,
Cohen et al. (1998) emphasize that approximately half of the galaxies
in their sample 
lie in five ``redshift peaks'', likely due to clustering. Therefore, Poisson
statistics may be more profitably applied to the number of sampled 
``structures'', rather than to that of galaxies. 

\subsection{The Role of Spheroids in the Cosmic History of Star Formation} 

 With spheroids containing at least 30\% of all
stars in the local universe (Schechter \& Dressler 1987; 
Persic \& Salucci  1992) or even more
(Fukujita, Hogan, \& Peebles 1998), one can conclude that at least
 30\% of all stars -- hence $\sim 30\%$ of metals -- 
have formed at $z\gsim 3$ (Renzini 1998a;
Dressler \& Gunn 1990).
This is several times more than suggested by a conservative
 interpretation of the early attempt at tracing the cosmic history of
 star formation, either empirically (Madau et al. 1996) or with  
theoretical simulations (e.g. Baugh et al. 1996). 
Yet, it is in fine agreement with the 
recent direct estimates from the spectroscopy of Lyman-break galaxies
(Steidel et al. 1998), as well as with sub-mm observations (Hughes et al.
1998), where the cosmic SFR runs flat for $z\gsim 1$,
as in one of the options offered by the models of Madau, Pozzetti, \&
Dickinson (1998). 

\subsection{The Role of Spheroids in the  Metal Enrichment of the
Early Universe}

The global metallicity of the present day universe is best estimated 
in clusters of galaxies, where it is  $\sim 1/3$ solar. This can be
taken as representative of the overall metallicity since
clusters and {\it field} have converted into star and galaxies nearly
the same fraction of baryons (Renzini 1997).
With $\sim 30\%$ of all stars having formed at $z\gsim 3$, and the
metallicity of the $z=0$ universe being $\sim 1/3$ solar, it is
straightforward
to conclude that the global metallicity of the $z=3$ universe had to
be at least $\sim 1/3\times 1/3\sim 1/10$ (Renzini 1998a,c). Damped Ly$_\alpha$
systems (DLA) may offer an opportunity to check this prediction, though they
may provide a vision of the early 
universe that is biased in favor of cold, metal-poor gas that has been
only marginally affected by star formation and metal pollution.
Metal rich objects that may exist at high redshift, such as giant
starbursts that would be dust obscured, 
metal rich passively evolving spheroids, and the hot ICM/IGM obviously do not
enlist among DLAs. Still, these objects may contain much of the metals in the
$z\sim 3$ universe as they do in the present day universe.
In spite of these limitations the average metallicity of the DLAs
at $z=3$ appears to be
$\sim 1/20$ solar (Pettini et al. 1997, see their Fig. 4), just a factor of
2  below the expected value from the {\it fossil evidence}. 
However, this is still much higher than the extreme lower limit 
$Z\simeq 10^{-3}Z_\odot$ at $z=3$ as
inferred from Ly$_\alpha$ forest observations (Songaila 1997). 
Ly$_\alpha$ forest material
is believed to contain a major fraction of cosmic baryons at high $z$, hence
(perhaps) of metals. There is therefore a potential conflict with
the estimated global metallicity at $z\simeq 3$, and the notion of 
Ly$_\alpha$ forest metallicity being representative of the the
universe metallicity at this redshift.
Scaling
down from the cluster yield, such low metallicity was achieved when
only $\sim 0.3\%$ of stars had formed, which may be largely
insufficient to
ionize the universe and keep it ionized up to this redshift (Madau
1998, but see Gnedin \& Ostriker 1997). 
This suggests that Ly$_\alpha$ forest may not trace the 
mass-averaged metallicity
of high redshift universe, and that the universe was very
inhomogeneous at that epoch. The bulk of metals would be partly locked
into stars in the young spheroidals, partly would reside 
in a yet undetected hot
IGM, a phase hotter than the Ly$_\alpha$ forest phase.

\subsection{Open Questions}
Several questions remain open at this stage. Some of them
can soon get answers from observations, others from new
theoretical simulations, or from extracting more information from old
ones. Of course, the list of interesting questions 
could actually be much longer, and include e.g. the origin(s) of all those
structural and morphological aspects that have been set deliberately
aside in this review.

\medskip
\spb
 How can hierarchical models be tuned to produce the uniform age of
  stars in the \par Galactic bulge? ...
\spb
 and the uniformity of stellar metallicity in the bulge of M31?
\spb What fraction of {\it ``ellipticals''} would belong to clusters,
 groups, and field in simulations \par of galaxy formation?
 \spb
 How much number evolution of spheroids has taken place between 
 $z=1$ and $z=0$?
\spb
 What is the redshift distribution of a fair and complete sample of $K\le 20$
 galaxies?
\spb
 Is the fraction of spheroids formed by merging {\it spirals} very
 large or very small?
\spb
 At which redshift do color-magnitude (and analogous) relations for
ellipticals begin to \par flatten? Do they flatten at all?
\spb
 At $z\simeq 1$ do global relations for ellipticals in the field differ
from those of galaxies in \par clusters, and if so by how much?
\spb
 Are Lyman-break galaxies spheroids in formation? What is their mass?
\spb
 What is the global metallicity of the universe at $z=3$?
\spb
 Does an early assembly of bulges help forming the right disks?
\spb
 Is the early universe re-ionized and maintained ionized by forming spheroids?
\smallskip
\par\noindent
It is my feeling that it will not take much before having fairly
secure answers to most of these questions.
\bigskip
\par\noindent
{\it Acknowledgements}: 
I would like to thank Ralf Bender, Marc Dickinson, Marijn Franx and 
Pascale Jablonka for their kind permission to reproduce here some of the
figures from their papers. I would also like to thank the Space
Telescope Science Institute for its hospitality during the meeting.

\end{document}